# Communities of Practice: Performance and Evolution

Bernardo A. Huberman and Tad Hogg

Dynamics of Computation Group
Xerox Palo Alto Research Center
Palo Alto, CA 94304

## Abstract

We present a detailed model of collaboration in communities of practice and we examine its dynamical consequences for the group as a whole. We establish the existence of a novel mechanism that allows the community to naturally adapt to growth, specialization, or changes in the environment without the need for central controls. This mechanism relies on the appearance of a dynamical instability that initates an exploration of novel interactions, eventually leading to higher performance for the community as a whole.

October 24, 1994

## Introduction

The existence of informal networks of collaboration within and across organizations is a well established and studied phenomenon[1]. Any institution that provides opportunities for communication among its members is eventually threaded by communities of people who have similar goals and a shared understanding of their activities [2]. These informal networks coexist with the formal structure of the organization and serve many purposes, such as resolving the conflicting goals of the institution to which they belong, solving problems in more efficient ways, and furthering the interest of their members, to name a few. In spite of their lack of official recognition, informal networks can provide effective ways of learning, a sharpened sense of belonging, and with the proper incentives actually enhance the productivity of the formal organization[2]. Equally important from a social point of view, whenever such informal networks appear they generate their own norms and interaction patterns, thus constituting what has been recently called a *community of practice* [3].

These communities, how they form and the roles they play, have been the focus of a number of studies over the years [4–6]. The research has ranged from the effectiveness of the invisible colleges in the progress of the scientific enterprise [7], to the roles of cliques in the functioning of bureaucracies [8]. In between, they run the gamut from informal networks of cooperation among chemists working for competitive pharmaceutical industries [9] to back channel exchanges between members of the foreign offices of adversary countries [10] and the appearance of gangs in schools [11] and prisons.

The pervasiveness of this phenomenon indicates that regardless of the nature of the embedding institution, there are general mechanisms that lead to the emergence of communities of practice, as well as to the evolution of their structures. One would like to find what these mechanisms are and how they depend on a number of variables, such as the size and diversity of the organization, the ability of its members to communicate, the nature of the problem being addressed, the incentives that make individuals join such communities and the structure and cost of the available communication media. As is well known, the natural limit to the number of individuals with which any member of the community can communicate with leads to interaction patterns that range from a flat type of structure, in which everybody interacts with everybody, to a clustered one in which individuals collaborate with a few others in the community [12]. Since these structures change in time due to many factors [13], one would like to know how do communities of practice evolve and how does the resulting structure depend on the overall size of the group and the diversity of skills available.

---

[1] The literature documenting this phenomenon is too extensive to be cited in detail. It is surveyed in [1]

[2] In many cases, informal networks can be detrimental to the organization they belong to. An early study that documents this negative influence within French culture is provided by Jesse R. Pitts, "In Search of France", Harvard University Press 1963.

We attempt to answer these questions by identifying some key elements responsible for cooperative problem solving by communities of practice, and by establishing how the members of such networks dynamically interact. An important aspect of our theory is the explicit realization that the performance of a single individual engaged in problem solving can often be enhanced by exchanging information with other members of its community. In spite of the fact that this information is not always perfect or even ideally suited for the task at hand, individuals learn to identify those other members of their community which are most useful to interact with. This results in an interaction pattern such that the frequency with which individuals interact with each other tends to scale with their mutual perceived usefulness. Since however, an individual cannot present novel hints to the solution of someone else's problems at every instant of time, other less useful individuals are then contacted when self work in itself cannot produce the required solution, thereby increasing the range of individuals contacted.

This picture of interactions is never static, for even when the nature of the problem does not change, those individuals able to make particular contributions tend to vary with time. This raises the interesting issue of the existence of any discernible pattern of interactions over time and their resilience to the unavoidable novelty that results from the casual encounters and discoveries that occur in the daily life of an organization. Whether such disturbances lead to a long term restructuring of the community, or to a transitory instability, depends on subtle dynamical effects which we elucidate.

In this paper, we present a detailed model of this collaborative performance enhancement and examine its dynamical consequences for the community as a whole. We do so by first establishing how the performance of a group of individuals depends on the rates at which they produce results that can be used by other members of the community, as well as the diversity and size of the group. When the links between members of the community change in time, we examine the resulting network structures and their dynamical stability against fluctuations. In order to make these results concrete we introduce an explicit learning mechanism that leads to local changes in the arrangements. We then show that these local changes lead to enhanced performance of the community as a whole, in agreement with empirical observations of the way work is organized in collaborative settings [14, 15].

On the dynamical side, we establish the existence of a novel mechanism that allows a community of practice to naturally adapt to growth, specialization, or changes in the environment, without the need for central controls. This mechanism is very general, and relies on the appearance of an instability in the structure of the community as it grows or becomes more diverse. This instability makes the community adaptable by taking it on a path whereby new interactions are explored so that it can eventually adapt to new goals and environmental changes.

This research complements the vast literature on informal networks that we mentioned above by the explicit introduction of performance measures and their dynamical dependence on interactions patterns that are allowed to take place in a very general fashion. On a methodological vein, we state that our main interest is in the average or typical behavior of communities, rather than detailed predictions of individual cases. We achieve this by focusing on those interactions that are ubiquitous in informal organizations. Ours is a bottom-up approach [16, 17] and it implies the introduction of probabilistic quantities that provide a coarse description of the myriad interactions observed in communities of practice. Although we are aware that this entails a loss of detailed specification, we believe that it is compensated by the predictive power of the formalism. On the other hand, an understanding of the idiosyncrasies of particular informal organizations will always require the more detailed studies that are part and parcel of the empirical tradition in organizational sociology.

## The Performance of a Community

Consider a group of individuals with diverse characteristics trying to solve a global problem. Individuals can interact with each other and if they choose to, they do so with an interaction strength proportional to the frequency with which they exchange information, or "hints", with each other. Since there is a natural limit, or bandwidth, to the number of people a given individual can interact with, the pattern of interactions ranges from the case in which individuals can interact with everybody very rarely, to the situation in which a member of the group interacts with few individuals very often.

A key issue for understanding the performance of a community of practice is how do its members make effective use of each others efforts, as well as their own. The simplest strategy would rely on some prespecified and fixed pattern. This strategy has the drawback of missing opportunities to exploit information gained about the problem and can also lack robustness against unanticipated changes [18]. Another common strategy is to use a global planning agent (e.g., a funding agency or management unit) to allocate resources to different groups and to adjust the organizational structure based on perceived progress. This is the way the scientific community is funded at the national level in most countries. While this can be responsive to unexpected changes and directly optimize some global performance measure, it becomes less feasible when the number of agents is large and the system is changing rapidly, since the planner may not be able to keep up with the pace of change and compute a response in a timely manner. More generally, the information needed to design the community may be distributed among the individuals and not readily available to any central planner. Finally, another strategy relies on a distributed control approach to the system that provides rapid and robust response and local reorganizations. While appealing, this approach can suffer from a lack of global perspective on the part of the individuals, thus leading to lower performance.

These considerations point to the need to relate the overall performance of a community of practice to its structure and the skills of the individuals within the organization. This relation can be obtained by first relating the overall performance of the group to that of its members, and then determining how the individual's performance depends on its skills and interactions with others in the organization.

There are many ways through which individuals can combine their efforts to produce something useful for a group [19]. A simple measure of their effectiveness would be provided by the sum of individual performances. It is appropriate for cases in which each individual performs a task that directly produces a single product or revenue for the community, in which case the performance is just the rate of production. It is less appropriate when there are conflicts among the individuals, e.g., due to resource contention or incompatible incentives [13], in which case a higher performance for one individual can imply a reduction in the performance of the others. Even in this case however, this measure gives the average individual performance, which may be of interest in its own right, as when evaluating the effect of policies such as training or potential reorganizations on the individuals involved. Quantitatively, we thus take the overall performance to be given by

$$P = \sum_i P_i \qquad (1)$$

where $P_i$ is the performance of individual $i$.

In our theory, we will assume that individuals proceed through a series of steps to complete their tasks. At each step, individuals can choose to work on their own (which we call "self-work") or use information or other help (which we call "hints") from others in the community. In some cases, during a given step a new hint will be produced that is made available to others in the organization. To quantify these choices, we characterize the informal structure of the organization by the pattern of interactions among its members, i.e., who talks to whom, and how frequently. Let $p_{ij}$ denote the probability that individual $i$ chooses to use a hint from agent $j$, and let $p_{ii}$ the probability that it instead performs self-work. Since the agents are active at each step the condition $\sum_j p_{ij} = 1$ holds. For simplicity, we assume that the steps are completed asynchronously at a rate $r$. Thus the rate at which $i$ uses a hint from $j$ is just $rp_{ij}$.

The final ingredient of our model of individual performance is the quality of the self-work and the effectiveness of the hints that are exchanged. Since we are interested in studying how the structure of the organization affects performance, we simply suppose all the self-work produces the same benefit, $s$, for each step. Thus the performance of an agent that does not use any hints (i.e., $p_{ij} = \delta_{ij}$ is 1 when $i = j$ and 0 otherwise) is simply given by $P_i = rs$.

To allow for the possibility of diversity within the community, we assume that the quality of the hints varies among the individuals. There are several contributing factors to this variability. First, the usefulness of a hint to an individual depends not only on the content of the hint itself, but also on how well it fits into the activities of the recipient. Second, if a hint is already known to an individual, it will be of little additional value, while it may be of great value to another individual without that information. Notice that useful hints not only improve the performance of the recipient, but also allow the recipient to produce better hints for others.

We model these effects as follows. First, let $h_{ij}$ be the quality of a new hint sent to *i* from *j* when *j* is receiving no hints (i.e., is doing self-work only). To focus on the situation in which cooperation is beneficial, we will generally consider the case in which most of the performance is due to hints, which amounts to take $h_{ij} > s$. As mentioned above, individual *i* will access such a hint from individual *j* at a rate given by $rp_{ij}$. If this rate is too high, successive hints will carry no novelty, resulting in a drop in their effective quality. If we assume that hints are produced at a rate, *w*, which is less than *r* (i.e., on average, an agent needs more than one step to produce a new hint), and that reusing a previously seen hint provides zero value, a simple approximate expression for the possibility of reusing an old hint gives an expected hint quality of

$$h_{ij}^{eff} \approx h_{ij}\left(1 - \frac{rp_{ij}}{w}\right) \qquad (2)$$

Combining these values then gives a simple value for individual performance in terms of the interaction structure of the community. It is given by:

$$P_i = r \sum_j p_{ij} h_{ij}^{eff} \qquad (3)$$

where, for simplicity, we have defined $h_{ii}^{eff} = s$.

| | |
|---|---|
| n | size of the organization |
| r | rate of completing steps |
| w | rate of producing hints, $w < r$ |
| $p_{ij}$ | organizational structure: probability *i* uses hint from *j* |
| s | quality of individual "self-work", $s \geq 0$ |
| $h_{ij}$ | quality of hints from *j* to *i*, $h_{ij} > s$ |

**Table 1.** Parameters used to describe an organization and its performance.

In spite of its simplicity, this model is somewhat more general than it may appear because a number of other effects can be considered by reinterpreting the variables used here. For instance, one can allow for some additional random variability in these values, in which case the result should be interpreted as the average performance. Also, there are other ways through which the use of hints could affect performance, e.g., by increasing the rate at which steps are completed. This can also be accounted by our model by allowing a larger number of hints to be produced per step, while keeping the overall rate fixed.

With these considerations we now investigate how the performance of a community of practice changes with its structure.

## Structure and Performance

In this section we investigate the question of what interaction structure gives the best performance as a function of the size and diversity of the informal organization. In this context, size is just the number of individuals composing the community, while diversity gives the range of hint qualities that they exchange

We first consider some simple cases. For instance, as mentioned above, if everyone acts independently ($p_{ij} = \delta_{ij}$) the performance is due to self-work, i.e. $P = nrs$. Another simple case is that of a flat community, i.e., one in which each member has equal links to all neighbors, so that $p_{ij} = \frac{1}{n}$, where $n$ is the size of the organization. In this case the individual performance will be given by

$$P_i = \frac{r}{n}\left[\left(1 - \frac{r}{nw}\right)(n-1)\bar{h}_i + s\right] \qquad (4)$$

where $\bar{h}_i = \frac{1}{n-1}\sum_{j \neq i} h_{ij}$ is the average quality of hints available to $i$. The overall performance is easily obtained by summing this expression over all the agents.

At the other extreme of the interaction pattern, each individual could choose to accept hints from only a single source, nominally the one with the highest hint quality. That is, $p_{ij} = \delta_{i_1 j}$ where the index $i_k$ denotes the individual giving the k[th] best hints for $i$. In this case, the individual performance is given by $P_i = r h_{i i_1}\left(1 - \frac{r}{w}\right)$. Notice that since individual $i$ is overusing the best hints, this results in a significantly lowered performance compared to working independently or in a flat interaction patterns.

A higher performing structure is obtained when each individual accepts hints more frequently from higher quality sources, but not so frequently that the quality drops significantly due to overuse. For a given set of parameters we can find the maximum performing organization (see appendix). An interesting question is then what happens to the structure of the community as its size and the diversity of hint qualities changes. One way to characterize its structure by a single parameter is to the count the number of neighbors an individual has, weighted by how frequently they interact. This measure implies that individuals have zero neighbors when they spend all their time doing self-work. Another possibility is to count the neighbors that individuals have, weighted by the probabilities of interaction, and normalized to the maximum number of links. This will range between 1, if an individual receives hints from a single source, and $n-1$ if individuals interact in a flat structured way, with equal contribution from all members of the group. We accordingly define

$$N_i = \begin{cases} 0 & \text{if } p_{ii} = 1 \\ \sum_{k=1}^{n-1} p_{ii_k}/p_{ii_1} \end{cases} \tag{5}$$

as the effective number of neighbors for individual $i$ in the community. Finally, to provide meaningful comparison among organizations of different sizes, we consider the clustering, which we define as the average value of the quantity $N_i/(n-1)$, averaged over all agents $i$.

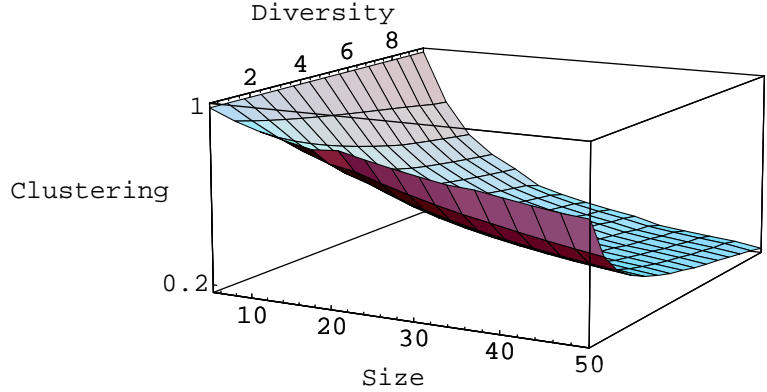

**Fig. 1.** Relative number of neighbors (clustering) for the optimal organization using uniformly spaced hints in $[10-\delta, 10+\delta]$ as a function of size $n$ and diversity $\delta$. The clustering ranges from 1 (i.e., a flat organization) to 0.19 (a fairly clustered organization). We use $r = 2.5$ and $w = 1$.

As a specific example, we consider the case in which the hint qualities are uniformly distributed within a range $[\mu - \delta, \mu + \delta]$. The resulting organizational structure is shown in Fig. 1. We see that small, uniform communities have best performance when their structure is fairly flat. As their size or diversity increases, it pays to specialize by concentrating only on the higher quality hints.

The behavior seen in this example applies more generally as well, and is due to a balance between an individual's capacity to use hints and the inevitable decrease in hint quality from overuse of a single source. Thus, in small homogeneous informal networks, individuals need to accept hints from most members in order to fill their capacity. With more diversity it eventually becomes beneficial to concentrate on the higher quality sources: even with a degradation in quality due to excessive use, the results obtained from them are still better than the much lower qualities produced by others. In addition, as a community gets larger with fixed diversity, there will be more relatively high quality sources to choose among. Thus an individual can concentrate on high quality hints from several sources and not overuse any single source.

This result is quite general, as it applies to other distributions of hint qualities. It explains why both increased size and diversity lead to more clustered organizational structures as being optimal. The implications of this finding are analogous to those of division of labor and specialization in the economics context.

## Evolution and Instabilities

We now turn to the dynamics of the interactions that characterize the communities of practice. This is an intriguing and important question since a group of individuals engaged in cooperative problem solving seldom settles into a fixed structure. Rather, the group is always characterized by interaction patterns that are constantly fluctuating. In light of these fluctuations, it is important to establish if there are indeed discernible patterns of interaction, how stable they are, and how they evolve. To this end, we study the general evolution of such networks and find the conditions that allow the community to achieve the optimal structure through purely local adjustments. This dynamical behavior of the informal organization (as defined by the evolution of the interaction patterns in the group) is to be contrasted to the more dramatic reorganizations imposed by formal organizations, which are defined by authority relations that in general do not incorporate all the information available locally to the individuals.

There are a variety of ways through which members of communities of practice change their interaction patterns. An important and familiar one is due to individual learning, which leads to improvement in self-work as well as to finding out who are the other members of the group that are good to listen to. This entails learning about the various hint qualities as well as how to make effective use of the hints, which in turn improves their effective quality. A large body of experimental evidence [20], along with theoretical studies of problem solving [21], shows that such learning can often be described by the power law in which the performance on a given task (e.g., self-work or the ability to effectively use a hint) improves as a power of the number of times the task is performed. With this learning mechanism the performance of a community of practice depends not only on its current structure but also on its history: the longer it has existed, the more time it has had to improve.

We are particularly interested in the extent to which local changes in the interaction patterns can improve the performance of the community, independent of improvements in individual performance. A simple way to study this is to allow the members of the community to increase their connectivity to those individuals that provide them with good hints. The natural limitation of the number of individuals with whom members can interact in a given time implies that this process leads to a decrease of their interactions with those now perceived to be less useful.

This picture leads to a general description of dynamics of the organization given by

$$\frac{dp_{ij}}{dt} = G_{ij}(\{p_{kl}\}) \tag{6}$$

where *G* is a function of all the probabilities which specify how individuals change their connections and the effectiveness with which they can use them because of learning of new options. We should note that since $\sum_j p_{ij} = 1$, the *G* function satisfies the condition $\sum_j G_{ij}(\{p_{kl}\}) = 0$. Since at this stage this function *G* is totally general, one expects that in order to elucidate the dynamics a detailed specification of its properties will be required. We will show however, that for a very large class of functions one can derive universal properties of the community dynamics, properties that tell of its stability and evolution as a function of its size and diversity. It is because of this universality that our results are of wide applicability and apply to a variety of individual learning mechanisms.

We start by noting that the solutions to this equation describe the evolving pattern of interactions in a community of practice. If these patterns ever equilibrate, one can find them by solving for the zeros of $G$. But finding the equilibrium patterns is not enough, since in any community there are always fluctuating interactions that disrupt the status quo. One can envision, for example, a situation whereby a casual conversation leads to the discovery that another member of the organization can contribute a partial answer to an ongoing problem, thus leading to a temporary change in the interactions between the individual and other members of the community. Since that change leaves other members with the possibility of interacting with each other with different frequencies, the small change precipitated by the encounter can cascade throughout the network and lead to a new structure. Alternatively, the network can recover from such a change and return to its original configuration.

This example illustrates the need to study the stability of the equilibrium patterns against small fluctuations. To see how this is done, suppose that a set of values for links, $p_{ij}$, corresponds to an equilibrium of Eq. 6. If there is a perturbation, $\epsilon_{ij}$ that takes the system away from the equilibrium by changing the instantaneous set of interactions, we investigate the ensuing dynamics by writing $p_{ij} + \epsilon_{ij}$ in Eq. 6. Since in equilibrium $\frac{dp_{ij}}{dt} = 0$ we obtain the following linearized matrix equation for the evolution of the perturbation:

$$\frac{d\epsilon}{dt} = M\epsilon \tag{7}$$

where *M* is the *Jacobian* matrix, which is given by

$$M_{kl} = \left(\frac{\partial G_{k_1 k_2}}{\partial p_{l_1 l_2}}\right) \tag{8}$$

evaluated at the equilibrium point. In this equation, $k$ and $l$ denote links between the specified individuals 1 and 2. The components of this matrix describe how a small increase in the frequency of interaction between a pair of individuals changes the interaction of another pair. Thus, the diagonal elements of *M* show the direct effect of a small change in the strength of an interaction between two individuals on the subsequent exchanges between them. In equilibrium one expects this effect to be such as to counteract the original change, which implies that the diagonal elements will be negative. On the other hand an off-diagonal entry describes the direct effect on a link from a change in another one, which can be of either sign.

The effect of the perturbation is determined by the eigenvalue of *M* with the largest real part, which we denote by *E*. Specifically, the long time behavior of the perturbation is given by $\epsilon_{ij} \propto e^{Et}$, implying that if *E* is negative, the disturbance will die away and the system return to its original interaction pattern. If, on the other hand, *E>0*, the smallest perturbation will render the community unstable.

The value of *E* depends on the particular choice of the interaction matrix. Methodologically, the study of the general behavior of interaction matrices is performed by examining the average properties of the class that satisfies all the known information about them. A class of plausible stability matrices is determined by the amount of information one has about particular learning mechanisms by individuals. For the sake of treating a very general case, we will assume as little knowledge about learning mechanisms as it is possible. This implies that all matrices that are possible Jacobians can be considered, and that there is no particular basis for choosing one over the other. This is the class of the so-called random matrices, in which all such matrices are equally likely to occur. Matrices in this class are such that each entry is obtained from a random distribution with a specified mean and variance.

In spite of their random nature, these matrices possess a number of well defined properties, among them the behavior of their eigenvalues [22, 23]. This means that we can use these properties to ascertain the stability of the network against perturbations in the interactions among individuals.

For example, when the nondiagonal terms have positive mean, $\mu$, standard deviation, $\sigma$, and the diagonal terms have mean, $\nu$, the largest eigenvalue grows with the size of the matrix as [24, 25]

$$E \sim \mu(n-1) + \nu \qquad (9)$$

Since in our case $\nu < 0$, this result implies that even if the community is stable when small (i.e. *E<0*), it will become unstable as its size becomes large enough for *E* to change sign. Furthermore even in the case when $\mu = 0$, the largest eigenvalue grows as $\sigma\sqrt{n}$. Even though this growth with size is much slower than the one we just considered, it grows linearly with diversity. This result again implies that as the community gets larger or more diverse, an instability against perturbations will always occur.

An even slower growth of the largest eigenvalue with size is obtained with more clustered interactions [26]. For example, tree structured matrices as might arise in a hierarchical pattern of interactions, have eigenvalues growing no faster than $\ln n$ implying that much larger communities can be stable when their members interact in a clustered fashion.

More complicated performance functions will arise if one takes into consideration indirect interactions among members of a community of practice. They are most likely to arise when two links share an individual. This is because when an individual spends more time interacting with a member of the network, the consequence will a reduction in the time spent with the other individual, or an increase if the indirect improvement in the hints is dominant. This leads to a non-diagonal matrix, whose non zero elements are only those for pairs of links that share a common individual. With *n* individuals there are $\binom{n}{2}$ links. For a given link, there are $2(n-2)$ other links that have a shared member with the given one. Therefore the fraction of non zero entries in the matrix is

$$\frac{2(n-2)}{\binom{n}{2}} \approx \frac{4}{n} \tag{10}$$

If the non zero elements of this matrix have mean *m*, and standard deviation *s*, then the mean of all the non diagonal terms will be $\mu = m\frac{4}{n}$. Similarly, the standard deviation for the non diagonal terms will be $\sigma = s\frac{4}{n}$. For this case, the previous discussion does not apply, so one has to resort to computer experiments to elucidate the properties of the eigenvalues of these matrices. These experiments show that they do not grow with the size of the system, implying stability of the community *independent* of its size.

This stability can be subverted if the diversity of the community, *s*, grows as the community gets larger. In this case, an instability will take place that is intermediate between the cases of $\mu = 0$ and $\mu > 0$ that we discussed above. In words, a growing community with an associated increasing diversity of interactions will not withstand the myriad perturbations that occur in its daily life without undergoing radical changes in its structure.

What can one conclude from these results? Typically, the interaction pattern represented by the entries in the Jacobian will correspond to the structure of the community given by the equilibrium point. Thus relatively flat structures, with a great deal of interaction, can be expected to have denser matrices than more clustered networks. As we just discussed, this means that flat informal networks become unstable more readily for given interaction strengths than clustered communities. This gives rise to a scenario in which flat interaction structures, suitable for small homogeneous communities, eventually become unstable and evolve towards a more robust equilibrium, characterized by a clustered community. Furthermore, this instability and the consequent growth of the more clustered network is accompanied by an increase in the overall performance of the community. This process of change is characterized by ever changing interaction frequencies, so that the evolving network can explore new possibilities for the solution of novel problems, thus becoming very adaptable.

Another dynamical issue has to do with variations in the size, diversity and environment of the community, as opposed to fluctuations in the interaction patterns while all other parameters remain fixed. Since in practice the environment in which the community is embedded tends to undergo changes, the appropriate equilibrium can also change in adaptive ways, and it is important to see under which conditions the community of practice can adjust accordingly. In the simplest case the new equilibrium moves smoothly to another configuration and there is no impediment for the internal dynamics to follow suit. For example as the size or diversity of the community grows, the equilibrium simply gradually shifts from flat to a clustered structure.

However, in more realistic situations there are various inertia effects that can prevent this smooth progression from a flat to a clustered community. Inertia can result from many organizational factors [27], transaction costs and, more fundamentally, the temporary loss in performance due to the need to relearn new connections as the community changes.

## Stability, Learning and Adaptation

Since the discussion of the previous section is very general, we now show how those results can be used to study communities of practice in more specific terms. In this section we look at a concrete example by selecting a simple case of individual learning and looking at the ensuing dynamics of the community as a whole. We will also consider the effects of learning on the performance of the group, and on its response to sudden changes in the environment. This is an important issue that also appears in the context of formal organizations [28]. In what follows we will consider the case when individuals increase their connections with those members of the group perceived to be most helpful to them. In other words, the interaction frequencies among individuals change in proportion to the performance that is obtained from having a link between them. In the this case the dynamics is determined by a *G* function that reads

$$G_{ij} = \alpha (\nabla P)_{ij} = \alpha \frac{\partial P}{\partial p_{ij}} \tag{11}$$

where $\alpha$ is the rate at which agents make changes in their connections. The performance *P* is given by Eq. 1 so the explicit dynamics is determined by Eq. 6. Rather than showing the analytical form of the solutions however, we now describe the ensuing behavior.

In this simplest case, the performance, *P*, will have a single maximum as a function of the structure of the community (which is determined by the links, $p_{ij}$). In this situation, the community turns out to be always stable. In the presence of fluctuations or even drastic changes in the structure of interactions, the informal network as a whole always moves towards its maximum performance. This adaptive behavior is illustrated in Fig. 2 and shows how this local dynamics can allow the organization to smoothly adjust to changes required as it grows or becomes more specialized. Note that in this figure the organizational structure, given by the values of the $p_{ij}$, is schematically illustrated by a single axis. In reality, there is a separate axis for each link, to represent how the performance depends on the strength of that individual link.

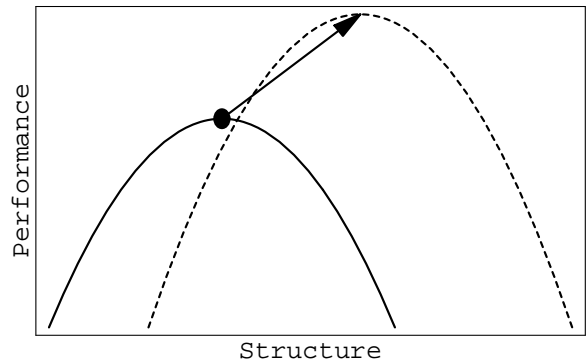

**Fig. 2.** Schematic behavior of the performance of the community as a function of its clustering. Along the horizontal axis, flat organizations are on the left, with increasing clustering toward the right. The solid curve shows the performance vs. structure for a small, homogeneous organization, with maximum at a fairly flat structure. The dashed curve is the performance for a larger or more specialized organization, with maximum corresponding to a more clustered organization. The arrow indicates how the original flat structure can become properly clustered by following the changing performance curve using local adaptations.

A more interesting case appears when learning enhances the individual performance based on how long the community has existed with a given interaction structure. This learning process leads to a functional form for the performance as a function of structure that is shown in the left hand panel of Fig. 3, which clearly shows the increased performance. The interesting situation from an adaptive point of view takes place when there is a sudden change in the environment in which the community is embedded, its size or diversity. For here one is interested in the ability of the whole community to respond in timely fashion to this sudden change. In terms of our dynamical model, this sudden change corresponds to a shift of the maximum of the performance function, as shown in the right hand panel of Fig. 3.

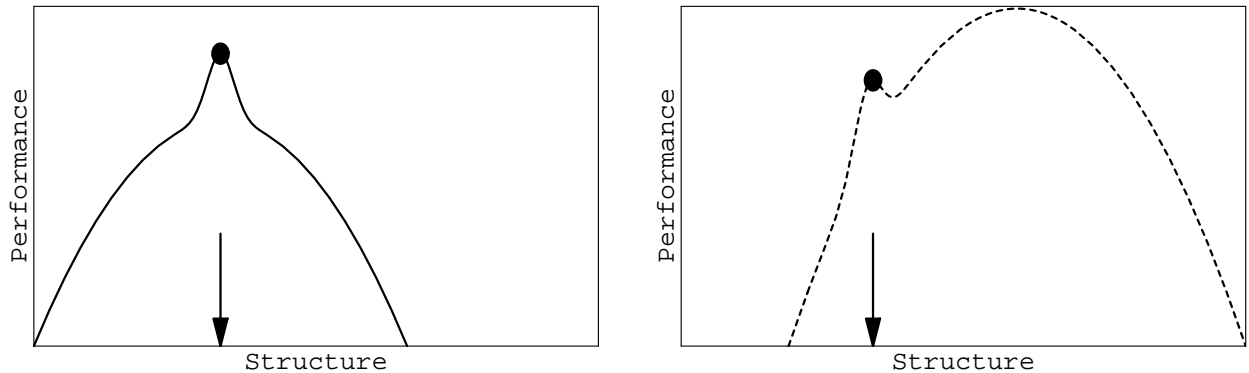

**Fig. 3.** The performance of the community when individuals learn. The arrows indicate the existing structure before any changes take place in the environment. Because of learning, the performance corresponding to this structure has improved. When the underlying performance changes, due to additional individuals, more diversity, or changes in the environment, the enhancement due to learning can prevent the organization from immediately adjusting.

Notice that unlike the previous case, a move away from the original performance maximum would now result in a temporary lowering of the performance of the community as a whole. Thus, this simple learning mechanism, in which individuals change in proportion to the performance that is obtained from having a link between them, will prevent the system from ever moving to the new maximal performance. Notice that in this case, while it may be tempting to use a central planner to remedy this non-optimal situation, such a strategy may not work well if the information available to each individual is not available to the central planner as well.

It is in this situation that the instabilities that we mentioned above play a crucial role in bringing the whole community into a new optimal configuration. As the community grows or becomes more diverse, the largest eigenvalue of the Jacobian becomes positive. The ensuing instability results in a novel pathway through which the performance of the community can grow to its maximal value without having to go through an intermediate decrease. As show in Fig. 4 the loss in stability can create a ridge (seen in the narrow darkened region) that circumvents the drop in performance shown in Fig. .3.

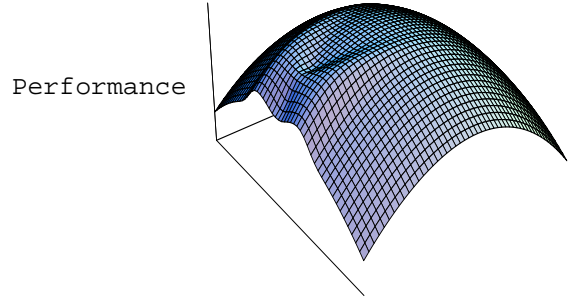

**Fig. 4.** Schematic behavior of performance as a function of structure (shown on two axes) for a community large enough or diverse enough to have undergone an instability. There is now a path, characterized by a set of evolving links, that can take the group to a higher performing point without ever lowering it.

Specifically, we can view the new axis as representing the direction, i.e., combination of changes in the $p_{ij}$ values, that corresponds to the unstable eigenvector of the Jacobian. While in other directions, represented by the original axis, the performance will be lowered by small changes, this unstable direction will allow the performance to smoothly increase. Hence the initial informal organization can move away from its original structure via the simple local dynamics of Eq. 11. This movement can continue to the new optimal point along the ridge shown in Fig. 4.

This accommodation to a set of new global constraints, along with its timely response, is what characterizes the adaptability of a complex system [29]. While this mechanism cannot guarantee a fast accommodation of the community of practice to very rapid environmental changes, it does provide a way out of the bottleneck that local learning can sometimes generate.

## Discussion

Since informal organizations are ubiquitous within institutional settings, it is of interest to evaluate their evolution and performance. This is especially timely in light of the recent decentralization trends exhibited by large corporations and government institutions. Along with these moves, there has been an increased tendency to empower individuals to solve problems in a timely fashion and to react quickly to perceived changes in the environment. This brings up the issue of the effectiveness of communities of practice, as well as their dynamics and adaptability to new environmental constraints.

In this paper we developed a theory of communities of practice that starting from the elementary interactions among their members and their actions, provides an understanding of some of their overall dynamical behaviors. In our theory members of the community can engage in self-work, receive hints from other members, and transmit their own findings for use by others. Within this framework, the links they form are seen to form a kind of ecosystem with interactions among them, competition for use, and imperfect decisions that are constantly being reevaluated. The theory allows for quantitative determinations of the performance of the community as a function of its shifting structure (i.e. who talks to whom) and it shows which patterns of interactions lead to optimal performance. At the dynamical level, we established a general mechanism for the changes in interactions that result from learning, and used it to establish the stability of a community of practice in light of the myriad disturbances and learning experiences that are part and parcel of the life of an organization.

An important new result is the existence of a dynamical instability in the pattern of interactions among members of the community. This instability takes place as either the diversity or the size of the organization increases, allowing the community to adjust to sudden changes in the environment. This provides an endogenous mechanism of adaptation that preempts the intervention of central controls. If there are no free riding problems in the community, these adaptive readjustments lead to better performance and increased utility to the individuals, in analogy to our findings in the context of organizational fluidity and sustainable cooperation [30].

A point worth stressing is that while the examples we used to illustrate our results are relatively simple, the results of the theory are quite general. This follows from the fact that our stability analysis does not depend on the specific functional form that one can choose for the performance of the community. While the precise numbers for which instabilities will set in depend on a number of organizational idiosyncrasies, the overall behavior will scale with size and diversity in the form that we predicted.

One may ask about the applicability of our predictions to real communities of practice. As with most theories of social systems, one is always confronted with the usual questions about their validation by social experiments [31]. Recent technological developments however, show promise in this direction, for they make it easier to observe informal interactions in an unobstrusive way. For example, the existence of vast electronic mail networks allows for studies of electronic mail use to throw light into the interaction patterns of the individuals. There are ways of providing anonymity that could make this an acceptable experimental tool.

On a different vein, our predictions also have normative implications. Since we have shown the ability of an informal network to display learning and adaptive behavior, one may question the need for global restructuring on the part of the formal organization. These reorganizations will always be needed when the transition to new forms is not proceeding at a satisfactory pace, and when some global coordination is needed. Alternatively, reorganizations might take place to align the formal organization with the informal one, thus leading to a more efficient output and consequent improvement in performance.

This point brings up the issue of incentives, for the existence of an optimally performing community of practice does not guarantee that its output and goals coincide with those of the embedding organization. How to align them is not only an aspect of management theory, but also an issue being faced by funding agencies and governments and interested in creating specific outputs [32]. This is because in the economy as a whole the firms fill the role of agents and their continual interactions produce an informal organization that continually adjusts to changes in markets and technology using local information [33].

Before closing we mention another instance of informal organizations whose evolution and performance is crucial for collective problem solving. The existence of computational ecologies [34] provides a natural framework for using these methods since they share a number of key features with human societies. These include asynchronous independent agents that solve problems from their local perspective involving uncertain and delayed information about the system. A number of attempts at collective problem solving from this perspective have been made and the resulting improvement in performance was measured [35–37]. When applied to this domain the results of our theory will allow for dynamically finding or evolving a better organization of the interaction patterns of multiagent systems, as well as a monitoring of their evolution.

As is well known, in human societies the benefit of cooperation underlies the existence of firms, exchange economies, scientific and professional communities, and the use of committees or teams charged with solving particular problems. Obtaining the full benefit of this potential improvement requires having the correct organizational structure to exchange relevant information and resources among the members of the community. Our study shows one way this can be achieved by a diverse and fluid group whose readjustments are the result of local behaviors. When coupled to the efficiency that results from having a formal embedding organization, our results show that the power of cooperative problem solving can be harnessed to deal with the most complex problems facing societies and institutions.

# Appendix A  Probability that a Hint is New

Consider an asynchronous process with rate $\alpha$. Then the probability that the next event (either a hint read or write in our case) occurs after time $t$ is $e^{-\alpha t}$. This follows from the assumption that events occur independently with probability $\alpha \Delta t$ in a small time interval $\Delta t$: if we take $n = t/\Delta t$, the probability for no event up to time $t$ is just $(1 - \alpha \Delta t)^n \to e^{-\alpha t}$ as we take $n \to \infty$.

Thus if we have two such processes, with rates $\alpha_1$ and $\alpha_2$, the probability that an event from the first process happens before one from the second is

$$\int_0^\infty dt \alpha_1 e^{-\alpha_1 t} e^{-\alpha_2 t} = \frac{1}{1 + \alpha_2/\alpha_1} \qquad (12)$$

In our application, the two processes correspond to $\alpha_1 = w$, the rate at which new hints are produced, and $\alpha_2 = r p_{ij}$, the rate at which they are used. With $h_{ij}$ being the value of a new hint, and assuming a previously used hint has zero value, the expected value of a hint with these rates will be $\frac{h_{ij}}{1+r p_{ij}/w}$. For simplicity, we expand the denominator to obtain a correction $h_{ij}\left(1 - \frac{r p_{ij}}{w}\right)$ which is quantitatively accurate provided the correction is small, and even for larger values gives the general qualitative behavior of decreasing expected hint quality when hints are used more frequently than they are produced.

# Appendix B  The Best Organizational Structure

Within the context of our model, we can determine the optimal organizational structure by maximizing the performance of each individual to determine the appropriate $p'_{ij}$s. Note that this is a constrained maximum in that the values must be between 0 and 1 and sum to 1.

Thus, we have to maximize

$$\begin{aligned} f &= \frac{1}{r} P_i = F + s\left(1 - \sum_{j \neq i} p_{ij}\right) \\ F &= \sum_{j \neq i} p_{ij} h_{ij}\left(1 - \frac{r p_{ij}}{w}\right) \end{aligned} \qquad (13)$$

The overall maximum satisfies

$$0 = \frac{\partial f}{\partial p_{ij}} = h_{ij}\left(1 - \frac{2r}{w} p_{ij}\right) - s \qquad (14)$$

giving the maximum at

$$p_{ij} = \frac{w}{2r}\left(1 - \frac{s}{h_{ij}}\right) \qquad (15)$$

This maximum corresponds to a possible organizational structure provided that the values are nonnegative and their sum does not exceed one, i.e.,

$$1 - \frac{2r}{w(n-1)} \leq \frac{s}{H_i} \tag{16}$$

where we have defined

$$\frac{1}{H_i} = \frac{1}{n-1} \sum_{j \neq i} \frac{1}{h_{ij}} \tag{17}$$

Note that $H_i > s$ and $p_{ij} > 0$ at the maximum point given above because we are supposing $h_{ij} > s$. Thus while this condition holds for small organizations, as $n$ increases with fixed hint qualities, it will eventually be violated.

For large organizations, it will no longer be optimal to perform any self-work, i.e., we will have $p_{ii} = 0$. In this case we need a Lagrange multiplier to find the maximum subject to this condition, i.e., maximize

$$f = F + \lambda \left(1 - \sum_{j \neq i} p_{ij}\right) \tag{18}$$

This gives

$$p_{ij} = \frac{w}{2r}\left(1 - \frac{\lambda}{h_{ij}}\right) \tag{19}$$

with $\lambda$ selected to make $\sum_{j \neq i} p_{ij} = 1$, i.e.,

$$\lambda = H_i \left(1 - \frac{2r}{w(n-1)}\right) \tag{20}$$

Note that this case applies only when $\lambda > s$, in particular requiring the multiplier to be nonnegative. (This is consistent: a negative value would imply more weight given to lower quality hints.)

However, this expression for the $p'_{ij}s$ can now give negative values when $n$ is sufficiently large that, e.g., $\lambda$ exceeds the smallest hint qualities. To see what happens then, consider the hints ordered according to their quality:

$$h_{ii_1} > h_{ii_2} > \ldots > h_{ii_{n-1}} > s \tag{21}$$

and suppose $m$ is the largest index such that $h_{ii_m} \geq \lambda > h_{ii_{m+1}}$ (note that it is always the case that $h_{ii_1} \geq \lambda$ since $\lambda < H_i \leq h_{ii_1}$). Then we must add additional constraints to the maximization:

$$f = F + \lambda \left(1 - \sum_{j \neq i} p_{ij}\right) + \sum_{k=m+1}^{n-1} \lambda_k p_{ii_k} \tag{22}$$

giving
$$p_{ii_k} = \frac{w}{2r}\left(1 - \frac{\lambda - \lambda_k}{h_{ii_k}}\right) \qquad (23)$$
with the definition $\lambda_k = 0$ for $k \leq m$, and the remaining $\lambda_k$ chosen to make $p_{ii_k} = 0$, i.e., $\lambda_k = \lambda - h_{ii_k}$. Finally the remaining probabilities must sum to 1, i.e., we get a new value
$$\lambda = H_{im}\left(1 - \frac{2r}{wm}\right) \qquad (24)$$
with
$$\frac{1}{H_{im}} = \frac{1}{m}\sum_{k=1}^{m}\frac{1}{h_{ii_k}} \qquad (25)$$
includes only the *m* largest hint qualities in the sum. Thus in this case only the top *m* hints are used.

In our example we took a uniform distribution of hint qualities. That is we picked $h_{ii_k} = \mu + \delta\left(1 - 2\frac{k-1}{n-2}\right)$ which ranges from a best hint quality of $h_{ii_1} = \mu + \delta$ to a worst of $h_{ii_{n-1}} = \mu - \delta$ which we take to be larger than *s*. For simplicity we set $s = 0$. Note that this gives the same distribution of hints to each agent that we used in Fig. 1.